\documentclass[
    ,final            
  ]
  {aipproc}

\layoutstyle{6x9}
\def\beq{\begin{equation}}
\def\eeq{\end{equation}}

\begin{document}

\rightline{McGill 02-22}
\bigskip
\title{Quest for a Self-Tuning Brane-World Solution to the Cosmological Constant Problem}

\author{James M. Cline}{
   address={Physics Department, McGill University, Montr\'eal, Qu\'ebec, Canada H2A 2T8}
,altaddress={presenter}
}

\author{Hassan Firouzjahi}{
address={Physics Department, McGill University, Montr\'eal, Qu\'ebec, Canada H2A 2T8}}

\begin{abstract}
 
It has been proposed that the geometry of an extra dimension could automatically adjust itself
to compensate for an arbitrary energy density on the 3-D brane which we are presumed to 
inhabit, such that a static solution to Einstein's equation results. This would solve the
long-standing cosmological constant problem, of why our universe is not overwhelmed by the
enormous energy of the quantum vacuum fluctuations predicted by quantum field theory. I will
review some of the attempts along these lines, and present a no-go theorem showing that these
attempts are doomed, at least within one of the most promising classes of models.

\end{abstract}

\maketitle


\section{The Cosmological Constant Problem}

One of the most vexing problems in theoretical particle physics is the magnitude
of the vacuum energy density, $\Lambda$.  Ironically, Einstein introduced it into general
relativity through the field equation
\beq
	G_{\mu\nu} \equiv R_{\mu\nu}-\frac12 g_{\mu\nu}R = 8\pi G(T_{\mu\nu} + \Lambda
g_{\mu\nu})
\eeq
in order to find static cosmological solutions.  He called it his ``biggest blunder''
when the universe was subsequently found to be expanding.  Generically the presence of
a positive cosmological constant leads to a universe which is accelerating, not static, but from the
Friedmann equation we see that the scale factor can be static for a fine-tuned value of 
$\Lambda$,
\beq
	\left({\dot a\over a}\right)^2 = {8\pi G\over 3}(\rho + \Lambda) - {k\over a^2}
\eeq
(in fact one must also tune $\ddot a$ to be zero).

The irony of Einstein's supposed blunder is three-fold. First, of course, the universe is not
static; second, particle theorists came to realize that $\Lambda$ should be present due to 
quantum fluctuations of the vacuum, which can be visualized as spontaneous creation and
annihilation of particle/antiparticle pairs.  Third, there is now strong evidence from
cosmology that $\Lambda$ is nonzero, 
\beq 
\Lambda \cong (2.4\times 10^{-3}\hbox{\ eV})^4 
\eeq
since the universe appears to be accelerating and at critical density, even though there is
not enough dark matter to account for most of the energy density.  The big problem is that
naive computations of the theoretical value of $\Lambda$ from quantum field theory give
a value which is many orders of magnitude greater,
\beq
	\Lambda_{\rm theo.} \sim (10^{28}\hbox{\ eV})^4
\eeq
There must be some mechanism for explaining the difference between the observed and expected
values, but so far no really convincing idea has been proposed.

\begin{figure}
\label{fig1}
  \includegraphics[height=.3\textheight]{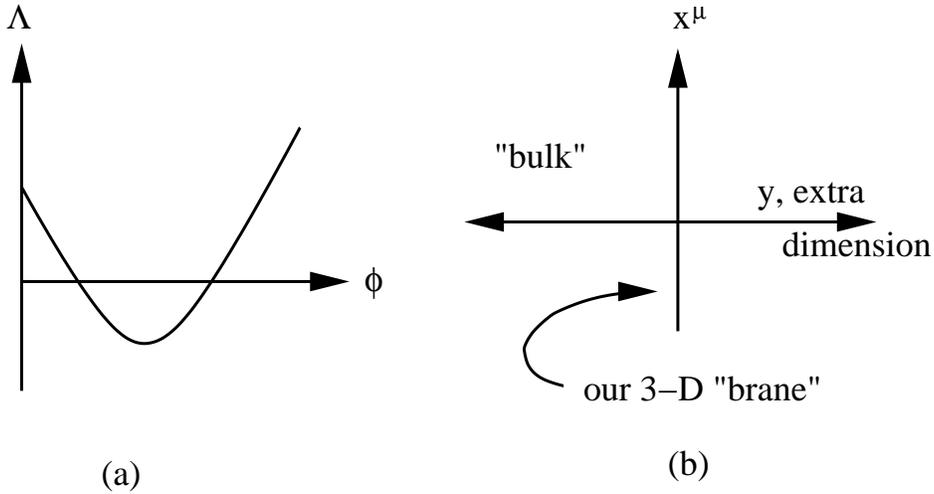}
  \caption{(a) A not very imaginative attempt at self-tuning.  (b)
The braneworld scenario.}
\end{figure}

One might suspect some kind of adjustment mechanism is at work, which somehow nullifies the
effect of $\Lambda$ no matter what its underlying value might be.  However Weinberg has given a
no-go theorem against such ideas \cite{Wein}.  For example, one might imagine that the effective physical
value of $\Lambda[\Lambda_0,\phi]$ depends on the underlying value $\Lambda_0$ and upon  a
scalar field $\phi$ which automatically adjusts itself so that $\Lambda[\Lambda_0,\phi]=0$.
To be concrete, consider $\Lambda = \Lambda_0 + f(\phi)$ as shown in fig.\ \ref{fig1}(a).
Although there are values of $\phi$ where $\Lambda=0$, they are not generally stationary,
so fine tuning is still required.  Although this is very obvious in the present example,
Weinberg's theorem shows that essentially the same problem will plague even more clever
attempts to implement such an idea.

Weinberg's theorem assumes that the universe is 4-dimensional, so one might hope that recent
brane-world ideas might provide a loophole.  In the braneworld picture illustrated in fig.\
1(b), we have an extra dimension, $y$, surrounding our 4-D universe which is presumed to be the
brane at $y=0$.  One could imagine that the effects of the vacuum energy density on the brane 
are counteracted by some new physics in the bulk.  If the 5-D nature of this setup can't be
described by an approximate 4-D picture, then it might be possible to evade the no-go theorem.

\section{Self-tuning solutions in 5-D}

Some attempts at constructing a self-tuning solution to the cosmological constant 
problem were presented in references \cite{ADKS, KSS}.  The idea is to add a scalar
field $\phi$ in the bulk, which is described by the metric 
\beq
	ds^2 = a^2(y) dx_\mu dx^\mu + dy^2
\eeq
The scalar is presumed to couple to the bare energy density of the brane, $\Lambda_0$ through a
potential $\Lambda_0 e^{-\kappa\phi}$.  The coupled field equations for the  scalar and the
metric give rise to a solution which is singular at some position $y_c$ in the bulk, as
illustrated in fig.\ 2.  The value of $y_c$ depends on $\Lambda_0$, which is what constitutes
the self tuning.  In other words, for any value of $\Lambda_0$, a static solution can be
found with some value of $y_c$.  It is precisely because the solution is static
rather than exponentially expanding that an observer
would infer that the physical value of $\Lambda$ is zero.
\begin{figure}
\label{fig2}
  \includegraphics[height=.25\textheight]{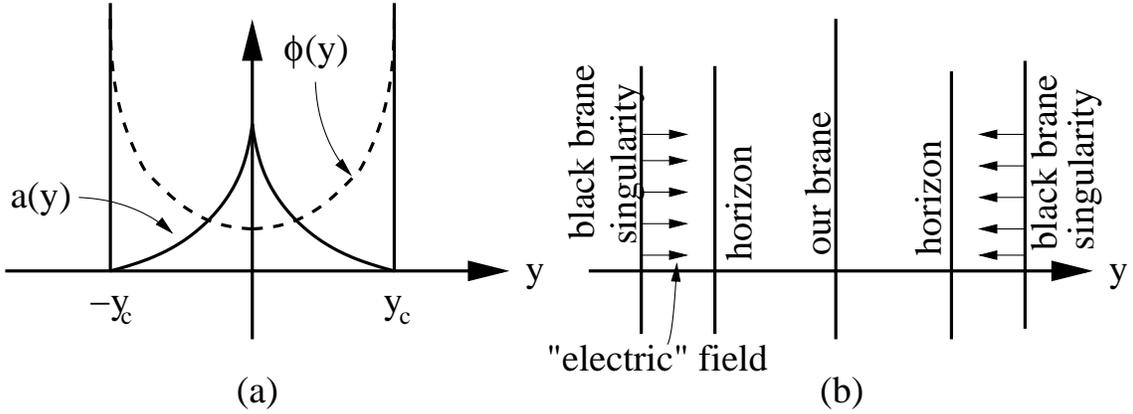}
  \caption{(a) Self-tuning brane solution.  (b) Brane-like singularities from the 
visible brane by horizons.}
\end{figure}

There are a number of problems with this idea \cite{Nilles}.  For one, the static solution is not  unique:
one can also find expanding or contracting ones \cite{BCG},  which shows that fine tuning of
the initial conditions are needed to get the static solution.  Moreover we don't like the
presence of naked singularities.  The cosmic censorship hypothesis asserts that singularities
should be hidden behind an event horizon, as is the case with a black hole.  Csaki {\it
et al.} \cite{CEG, CEG2} have proposed a modification to the previous self-tuning model which includes a
horizon, as shown in fig.\ 2(b).  The points $y\leftrightarrow -y$ are identified with 
each other by imposing a $Z_2$ orbifolding.

In this solution, the metric becomes the AdS-Reissner-N\"ordstrom solution shown in
fig.\ 3, 
\beq
	ds^2 = -h(y) dt^2 + a^2(y) d{\vec x}^2 + h^{-1}(y) dy^2
\eeq
with $h(y) = y^2/l^2 - \mu/y^2 + Q^2/y^6$ and $a^2(y) = y^2$; $l$ is the radius of
curvature of the 5-D anti-deSitter space, $\mu$ is mass of the black brane and
$Q$ is its charge under a U(1) gauge symmetry.  The self-tuning is now accomplished by
$\mu$ and $Q$ taking on the appropriate values to cancel the effect of the energy density
$\rho$ which is on the brane.  Note that these quantities appear not as inputs to the
Lagrangian, but rather as integration constants in the bulk solution.
However, this $\rho$ cannot be quite the same thing as a 
cosmological constant because the self-tuning solution works only for a rather bizarre equation of
state for the energy density on the brane:
\beq
	p < -\rho
\eeq
This is in contrast to a vacuum energy density which obeys $p=-\rho=-\Lambda$, and is in
violation of the weak energy condition.  Such violations may not always be bad, but at least
in the case of a classical scalar field theory they are problematic. There we
have  $\rho = \frac12\dot\phi^2 + V(\phi)$ and $p = \frac12\dot\phi^2 - V(\phi)$ so that
$p=-\rho$ implies $\dot\phi^2<0$: negative kinetic energy.  This would correspond to a 
ghost which leads to nonconservation of probability in quantum field theory.  Our 
goal in the present work was to try to find a self-tuning solution with a horizon and
without violating the weak energy condition.  Unfortunately, we found instead another
no-go theorem: one must have $p<-\rho$ either on the brane or in the bulk, as we will now
describe \cite{CF}.

\begin{figure}
\label{fig3}
  \includegraphics[height=.25\textheight]{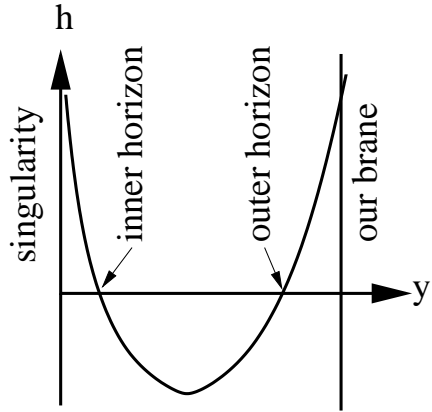}
  \caption{Metric function $h(r)$ for the 5-D AdS-Reissner-N\"ordstrom spacetime.}
\end{figure}

\subsection{No-go theorem}

Before discovering our no-go result, we searched for a generalization of the original
solution which would hopefully give us more leeway.  A natural possibility is to
try adding a scalar field in the bulk, so the Lagrangian becomes
\beq
	{\cal L} = \sqrt{|g|}\left(\frac12 g^{\mu\nu}\partial_\mu\phi\partial_\nu\phi
	-V(\phi) - V_0(\phi)\,{\delta(y)\over\sqrt{g_{55}}}\right)
\eeq
We obtained numerical solutions using various potentials and found empirically that a 
horizon could be obtained only when the black hole charge $Q^2$ became negative.  Interestingly,
$Q^2<0$ implies that $p<-\rho$ for the bulk stress energy tensor contributed by the gauge
field.

It is not difficult to show that the violation of the weak energy condition is 
actually necessary for obtaining a horizon.  By adding together the (00) and one of the
spatial ($ii$) components of the Einstein equations, and integrating from the outer
horizon ($y=y_h$) to the brane ($y=y_b$),  we obtain
\beq
\label{nogo}
	h'(y_h) = - {\kappa^2\over a^3}\left(2\int_{y_h}^{y_b} (T^0_{\ 0}-T^1_{\ 1})
	a^3 dy + \sqrt{h}(\rho+p)|_{y_b} \right)
\eeq
where $\kappa^2$ is the 5-D gravitational constant and $T^\mu_{\ \nu}$ is the bulk
stress energy tensor.  In particular, $(T^0_{\ 0}-T^1_{\ 1})$ is the bulk contribution
to $(\rho + p)$.  Notice the minus sign in front of everything, and the fact that $h'(y_h)$
must be positive at the outer horizon (see fig.\ 3).  These two facts conspire to need
$\rho+p$ to be negative somewhere, in the bulk, the brane or both.  

This conclusion is
independent of any details of the theory like the choice of scalar potentials.  We have
also found it to be robust against other generalizations: arbitrary field content in the bulk,
addition of higher derivative terms to the gravitational action, such as Gauss-Bonnet, and
also relaxation of the $Z_2$ symmetry we assumed around the brane.  For example, the
GB action is
\beq
 S = {1\over 2\kappa^2}\int d^{\,5}x\sqrt{|g|}\left(R + \lambda(R^2 - 4R_{ab}R^{ab} 
+ R_{abcd}R^{abcd}) \right)
\eeq
We find that the no-go result (\ref{nogo}) gets modified by the factor
\beq
\left.\left(1-\lambda (a'/a)^2 h\right) h' \right|_{y_h} = \hbox{same as in (\ref{nogo})}
\eeq	
But since $h=0$ at the horizon by definition, the left-hand-side is in effect unmodified.

Relaxing the $Z_2$ symmetry amounts to having two different black holes with different values of
$\mu$ and $Q$ on the two sides of the branes, with horizons at different distances.  This does
not change the conclusion.

There is one apparent loophole, which is to allow for spatial curvature on the brane.  The
Friedmann equation becomes $H^2 = (8\pi G/3)(\rho+\Lambda) - k/a^2$ in 4-D, where
$k=1,\ 0,\ -1$ for positive, zero or negative spatial curvature, respectively.  In our no-go
theorem we assumed $k=0$, which is consistent with observations of the cosmic microwave 
background and expectations from inflation.  With this modification we obtain
\beq
h'({y_h}) = \hbox{same as in (\ref{nogo})} + {4k\over a^3}\int_{y_h}^{y_b} a\, dy
\eeq
This term allows us to obtain a positive value for $h'({y_h})$ even if $\rho+p\ge 0$ 
everywhere, provided that $k>0$.  However, this does not solve the cosmological constant
problem because it requires that $\Lambda_0$ already be of order $k/Ga^2\sim (10^{-4}$ 
eV$)^4$, which is the original fine-tuning problem all over again.  Larger values of
$\Lambda_0$ would require larger curvatures, which are observationally excluded.

As a final attempt, one might try the same trick except with a very large curvature in
some hidden extra dimensions rather than the large ones.  However, the new terms in the
Einstein equations which depend on the extra-dimensional curvature exactly cancel out of 
the combination $G_{00} + G_{ii}$ which gave us the no-go result, so this does not help.

\section{Conclusion}

It seems that Nature abhors the self-tuning idea.  Each time we attempt to cure one  problem,
it introduces a new pathology.  Ref.\cite{CEG} tried to cure the problem of the naked
singularity by hiding it behind a horizon, but had to violate the weak energy  condition to do
so.  We find that allowing for positive spatial curvature can cure this problem, but only at
the price of reintroducing the original fine-tuning problem.  It is a good illustration of why
the cosmological constant problem is one of the most daunting in theoretical physics.  Some
progress with self-tuning seems to have been made recently in ref.\cite{BCDD}.  However, it
remains a challenge to understand how constants of integration which correspond to physical
sources of stress energy can dynamically adjust in a situation where the vacuum energy is
changing, as in a phase transition, or alternatively, how they can ``know'' which value is the
ultimate one that must be canceled.  It would seem to be a very intelligent mechanism that
could fully allow for the effects of vacuum energy during inflation but not in the present.

\begin{theacknowledgments}
Our work is supported by the Natural Sciences and Engineering Research Council of 
Canada and FCAR of Qu\'ebec. 
\end{theacknowledgments}





\end{document}